\pdfoutput=1

\documentclass[english]{article}
\usepackage[latin9]{inputenc}
\usepackage{geometry}
\usepackage{amsmath}
\usepackage{esint}

\makeatletter
\usepackage{hyperref}

\usepackage{mathtools}

\DeclarePairedDelimiterX\braket[2]{\langle}{\rangle}{#1 \delimsize\vert #2}

\date{}

\renewcommand{\vec}{\boldsymbol}

\makeatother

\usepackage{babel}
\begin{document}

\title{Far-field radiation of electric and toroidal dipoles in loss-less
non-magnetic dielectric medium with refractive index $n$}

\author{V. Savinov\thanks{Optoelectronics Research Centre and Centre for Photonic Metamaterials,
University of Southampton, Southampton SO17 1BJ}}
\maketitle
\begin{abstract}
Using simple classical treatment it is shown that the power emitted
by the electric and toroidal dipoles scales differently with the refractive
index of the surrounding environment.
\end{abstract}

\section{Introduction}

Toroidal dipole is an elementary electromagnetic excitation that can
be visualized as currents flowing along minor loops of an infinitesimal
torus (poloidal currents) \cite{Afa95}. The radiation pattern of
a dynamic toroidal dipole is identical to that of a dynamic electric
dipole - a pair of opposite (oscillating) charges \cite{Afa95}. 

The experimental exploration of the physical properties of toroidal
excitations became possible only recently, with the advent of electromagnetic
metamaterials, and is now attracting considerable attention\cite{Papasimakis16}.

The aim of this short note is to demonstrate that whilst the radiation
pattern of electric and toroidal dipoles may be identical, the power
emitted by these two excitations does scale differently with the ambient
refractive index. This question has already been explored in the context
of spontaneous decay rates of mesoscopic sources \cite{Tkalya02}.
Here we offer a much simpler exposition for point-like classical sources. 

\section{Power radiated by a point-like electric dipole embedded into an isotropic
dielectric medium}

The current density of a point-like electric dipole with moment $\vec{p}$,
located at the origin, is given by \cite{Afa95}:
\[
\vec{J}_{p}=\frac{d\vec{p}}{dt}\delta^{\left(3\right)}\left(\vec{r}\right)=\dot{\vec{p}}\delta^{\left(3\right)}\left(\vec{r}\right)
\]

In case of time-harmonic current (angular frequency $\omega$; assumed
time-dependence $\exp\left(i\omega t\right)$), the vector potential
due to $\vec{J}_{p}=i\omega\vec{p}\delta^{\left(3\right)}\left(\vec{r}\right)$
is\footnote{Hereinafter we shall only consider the fields away from the source.
The source term, i.e. a delta-function, that appears in the complete
solution for the vector potential is therefore omitted.} \cite{jackson}:
\[
\vec{A}_{p}=\frac{\mu_{0}}{4\pi}\int d^{3}r'\frac{\exp\left(-ik_{0}n\left|\vec{r}-\vec{r}'\right|\right)}{\left|\vec{r}-\vec{r}'\right|}\vec{J}_{p}=\frac{i\omega\mu_{0}}{4\pi}\,\frac{\exp\left(-ik_{0}nr\right)}{r}\,\vec{p}
\]

Where $k_{0}=\omega/c$ is the free-space wave-number, $n$ is the
refractive index of the medium, $c$ is the speed of light, and $\mu_{0}$
is the vacuum permeability. The far-field electric ($\vec{E}_{p}$),
and magnetic ($\vec{B}_{p}$) fields are given by:
\begin{flalign*}
\lim_{r\to\infty}\vec{B}_{p}= & \lim_{r\to\infty}\vec{\nabla}\times\vec{A}_{p}=\frac{i\omega\mu_{0}}{4\pi}\lim_{r\to\infty}\vec{\nabla}\times\frac{\exp\left(-ik_{0}nr\right)}{r}\,\vec{p}\\
= & \frac{i\omega\mu_{0}}{4\pi}\left(-ik_{0}n\vec{\hat{r}}\right)\times\frac{\exp\left(-ik_{0}nr\right)}{r}\,\vec{p}\\
= & \frac{\omega^{2}\mu_{0}n}{4\pi c}\,\left(\vec{\hat{r}}\times\vec{p}\right)\,\frac{\exp\left(-ik_{0}nr\right)}{r}\\
\lim_{r\to\infty}\vec{E}_{p}= & \lim_{r\to\infty}\frac{c^{2}}{i\omega n^{2}}\vec{\nabla}\times\vec{B}_{p}=\frac{c^{2}}{i\omega n^{2}}\lim_{r\to\infty}\vec{\nabla}\times\vec{\nabla}\times\vec{A}_{p}\\
= & \frac{c^{2}\mu_{0}}{4\pi n^{2}}\lim_{r\to\infty}\vec{\nabla}\times\vec{\nabla}\times\frac{\exp\left(-ik_{0}nr\right)}{r}\,\vec{p}\\
= & \frac{c^{2}\mu_{0}}{4\pi n^{2}}\left(-ik_{0}n\vec{\hat{r}}\right)\times\left(-ik_{0}n\vec{\hat{r}}\right)\times\frac{\exp\left(-ik_{0}nr\right)}{r}\,\vec{p}\\
= & \frac{-\mu_{0}\omega^{2}}{4\pi}\left(\vec{\hat{r}}\times\vec{\hat{r}}\times\vec{p}\right)\,\frac{\exp\left(-ik_{0}nr\right)}{r}
\end{flalign*}

The radiated power is given by the integral of the radial component
of the time-averaged Poynting vector ($\langle\vec{S}\rangle=\frac{1}{2\mu_{0}}\Re\left(\vec{E}\times\vec{B}^{\dagger}\right)$)
over the surface of an origin-centered sphere with radius $R\to\infty$:
\begin{flalign*}
P_{p}= & \lim_{R\to\infty}\oint_{R}d^{2}r\,\vec{\hat{r}}.\left(\frac{1}{2\mu_{0}}\Re\left(\vec{E}_{p}\times\vec{B}_{p}^{\dagger}\right)\right)=\frac{1}{2\mu_{0}}\,\frac{-\mu_{0}\omega^{2}}{4\pi}\,\frac{\omega^{2}\mu_{0}n}{4\pi c}\int d^{2}\Omega\,\vec{\hat{r}}.\left(\left(\vec{\hat{r}}\times\vec{\hat{r}}\times\vec{p}\right)\times\left(\vec{\hat{r}}\times\vec{p}^{\dagger}\right)\right)\\
P_{p}= & \frac{-\omega^{4}\mu_{0}n}{32\pi^{2}c}\,\left(\frac{-8\pi}{3}\left|\vec{p}\right|^{2}\right)=\frac{\mu_{0}\omega^{4}}{12\pi c}\cdot n\cdot\left|\vec{p}\right|^{2}
\end{flalign*}

Thus the power radiated by the electric dipole is linearly proportional
to the refractive index of the medium ($n$).

\section{Power radiated by a point-like toroidal dipole embedded into an isotropic
dielectric medium}

The current density of a point-like toroidal dipole with moment $\vec{T}$,
located at the origin, is given by \cite{Afa95}:
\[
\vec{J}_{T}=\vec{\nabla}\times\vec{\nabla}\times c\vec{T}\delta^{\left(3\right)}\left(\vec{r}\right)
\]

The corresponding vector potential is\footnote{One can invoke integration by parts to transfer the derivatives from
delta function to Green function, e.~g. $\int d^{3}r'\,G\left(\vec{r}-\vec{r}'\right)\vec{\nabla}'\times\vec{M}\left(\vec{r}'\right)=-\int d^{3}r\,\vec{\nabla'}\times G\left(\vec{r}-\vec{r}'\right)\vec{M}\left(\vec{r}'\right)=\vec{\nabla}\times\int d^{3}r\,G\left(\vec{r}-\vec{r}'\right)\vec{M}\left(\vec{r}'\right)$
if $\vec{M}$ vanishes at the boundaries.}:
\[
\vec{A}_{T}=\frac{\mu_{0}}{4\pi}\int d^{3}r'\frac{\exp\left(-ik_{0}n\left|\vec{r}-\vec{r}'\right|\right)}{\left|\vec{r}-\vec{r}'\right|}\vec{J}_{T}=\frac{\mu_{0}}{4\pi}\,\vec{\nabla}\times\vec{\nabla}\times c\vec{T}\frac{\exp\left(-ik_{0}nr\right)}{r}
\]

The far-field electric ($\vec{E}_{T}$), and magnetic ($\vec{B}_{T}$)
fields are given by\footnote{Since for toroidal dipole the charge density is $\rho_{T}=\vec{\nabla}.\vec{J}_{T}/\left(-i\omega\right)=0$,
one can use $\vec{E}_{T}=-i\omega\vec{A}_{T}$}:
\begin{flalign*}
\lim_{r\to\infty}\vec{B}_{T}= & \lim_{r\to\infty}\vec{\nabla}\times\vec{A}_{T}=\frac{\mu_{0}}{4\pi}\lim_{r\to\infty}\vec{\nabla}\times\vec{\nabla}\times\vec{\nabla}\times\frac{\exp\left(-ik_{0}nr\right)}{r}\,c\vec{T}\\
= & \frac{\mu_{0}}{4\pi}\left(-ik_{0}n\vec{\hat{r}}\right)\times\left(-ik_{0}n\vec{\hat{r}}\right)\times\left(-ik_{0}n\vec{\hat{r}}\right)\times\frac{\exp\left(-ik_{0}nr\right)}{r}\,c\vec{T}\\
= & i\frac{\omega^{3}\mu_{0}n^{3}}{4\pi c^{3}}\,\left(\vec{\hat{r}}\times\vec{\hat{r}}\times\vec{\hat{r}}\times c\vec{T}\right)\,\frac{\exp\left(-ik_{0}nr\right)}{r}\\
\lim_{r\to\infty}\vec{E}_{T}= & \lim_{r\to\infty}-i\omega\vec{A}_{T}=\frac{-i\omega\mu_{0}}{4\pi}\lim_{r\to\infty}\vec{\nabla}\times\vec{\nabla}\times\frac{\exp\left(-ik_{0}nr\right)}{r}\,c\vec{T}\\
= & \frac{-i\omega\mu_{0}}{4\pi}\left(-ik_{0}n\vec{\hat{r}}\right)\times\left(-ik_{0}n\vec{\hat{r}}\right)\times\frac{\exp\left(-ik_{0}nr\right)}{r}\,c\vec{T}\\
= & i\frac{\mu_{0}\omega^{3}n^{2}}{4\pi c^{2}}\left(\vec{\hat{r}}\times\vec{\hat{r}}\times c\vec{T}\right)\,\frac{\exp\left(-ik_{0}nr\right)}{r}
\end{flalign*}

The radiated power:
\begin{flalign*}
P_{T}= & \lim_{R\to\infty}\oint_{R}d^{2}r\,\vec{\hat{r}}.\left(\frac{1}{2\mu_{0}}\Re\left(\vec{E}_{T}\times\vec{B}_{T}^{\dagger}\right)\right)\\
= & \frac{1}{2\mu_{0}}\,\frac{\mu_{0}\omega^{3}n^{2}}{4\pi c^{2}}\,\frac{\omega^{3}\mu_{0}n^{3}}{4\pi c^{3}}\int d^{2}\Omega\,\vec{\hat{r}}.\left(\left(\vec{\hat{r}}\times\vec{\hat{r}}\times c\vec{T}\right)\times\left(\vec{\hat{r}}\times\vec{\hat{r}}\times\vec{\hat{r}}\times c\vec{T}^{\dagger}\right)\right)\\
P_{T}= & \frac{\omega^{6}\mu_{0}n^{5}}{32\pi^{2}c^{5}}\,\left(\frac{8\pi}{3}\left|c\vec{T}\right|^{2}\right)=\frac{\mu_{0}\omega^{6}}{12\pi c^{3}}\cdot n^{5}\cdot\left|\vec{T}\right|^{2}
\end{flalign*}

Thus the power radiated by the toroidal dipole is proportional to
the fifth power of the refractive index of the medium.

\section{Conclusion}

It has been demonstrated that the power emitted by a point-like electric
dipole scales linearly with the refractive index of the ambient environment,
whereas the emission of a point-like toroidal dipole scales as the
fifth power of the ambient refractive index.



\begin{thebibliography}{1}
\providecommand{\url}[1]{#1}
\csname url@samestyle\endcsname
\providecommand{\newblock}{\relax}
\providecommand{\bibinfo}[2]{#2}
\providecommand{\BIBentrySTDinterwordspacing}{\spaceskip=0pt\relax}
\providecommand{\BIBentryALTinterwordstretchfactor}{4}
\providecommand{\BIBentryALTinterwordspacing}{\spaceskip=\fontdimen2\font plus
\BIBentryALTinterwordstretchfactor\fontdimen3\font minus
  \fontdimen4\font\relax}
\providecommand{\BIBforeignlanguage}[2]{{%
\expandafter\ifx\csname l@#1\endcsname\relax
\typeout{** WARNING: IEEEtran.bst: No hyphenation pattern has been}%
\typeout{** loaded for the language `#1'. Using the pattern for}%
\typeout{** the default language instead.}%
\else
\language=\csname l@#1\endcsname
\fi
#2}}
\providecommand{\BIBdecl}{\relax}
\BIBdecl

\bibitem{Afa95}
{G. N. Afanasiev and Yu. P. Stepanovsky}, ``{The electromagnetic field of
  elementary time-dependent toroidal sources},'' \emph{J. Appl. Phys. A},
  vol.~28, p. 4565, 1995.

\bibitem{Papasimakis16}
{N. Papasimakis, V. A. Fedotov, V. Savinov, T. A. Raybould, and N. I.
  Zheludev}, ``{Electromagnetic toroidal excitations in matter and free
  space},'' \emph{Nature Mater.}, vol.~15, p. 263, 2016.

\bibitem{Tkalya02}
{E. V. Tkalya}, ``{Spontaneous electric multipole emission in a condensed
  medium and toroidal moments},'' \emph{Phys. Rev. A}, vol.~65, p. 022504,
  2002.

\bibitem{jackson}
{J. D. Jackson}, \emph{{Classical Electrodynamics}}.\hskip 1em plus 0.5em minus
  0.4em\relax {Wiley, New York ed. 3}, 1999.

\end{thebibliography}
\end{document}